\documentclass[11pt,a4paper]{article}
\pdfoutput=1

\usepackage{amsmath}
\usepackage{amssymb}
\usepackage{cite}
\usepackage{graphicx}
\usepackage{bbm}

\newcommand{\be}{\begin{equation}}  
\newcommand{\ee}{\end{equation}}

\newcommand{\vev}[1]{\langle #1 \rangle}

\newcommand{\SU}[1]{\ensuremath{\mathrm{SU}(#1)}}
\newcommand{\U}[1]{\ensuremath{\mathrm{U}(#1)}}
\newcommand{\into}{\ensuremath{\,\rightarrow\,}}
\newcommand{\tr}{\operatorname{tr}}

\newcommand{\Tb}{\ensuremath{\mathbf{T}}}
\newcommand{\T}{\ensuremath{\mathrm{T}}}
\newcommand{\wt}[1]{\widetilde{#1}}
\newcommand{\SUF}{\ensuremath{\mathrm{SU}(3)_{\rm F}}}
\newcommand{\gF}{\ensuremath{g_{\rm F}}}
\newcommand{\epsK}{\ensuremath{\varepsilon_K}}
\newcommand{\LamF}{\ensuremath{\Lambda_{\rm F}}}
\renewcommand{\to}{\ensuremath{\text{ to }}}

\begin{document}

\begin{titlepage}

\begin{flushright}
DESY 13-152\\
CERN-PH-TH/2013-291\\
WITS-CTP-124
\end{flushright}

\vskip 1.35cm
\begin{center}
{\bf\Large 
Light third-generation squarks from flavour gauge messengers
}

\vskip 1.2cm
Felix Br\"ummer$^{a,b}$, Moritz McGarrie$^{b,c}$ and Andreas Weiler$^{b,d}$
\vskip 0.4cm
{\it $^{a}$~SISSA/ISAS, I-34136 Trieste, Italy\\
$^{b}$~Deutsches Elektronen-Synchrotron DESY, D-22603 Hamburg, Germany\\
$^{c}$~National Institute for Theoretical Physics, School of Physics, and Centre for Theoretical Physics,
University of the Witwatersrand, Johannesburg, WITS 2050, South Africa\\
$^{d}$~CERN, Theory Division, Geneva, Switzerland}
\vskip 1.5cm

\abstract{\noindent
We study models of gauge-mediated supersymmetry breaking with a gauged horizontal $\SUF$ symmetry acting on the quark superfields. If $\SUF$ is broken non-supersymmetrically by $F$-term vacuum expectation values, the massive gauge bosons and gauginos become messengers for SUSY breaking mediation. These gauge messenger fields induce a flavour-dependent, negative contribution to the soft masses of the squarks at one loop. In combination with the soft terms from standard gauge mediation, one obtains large and degenerate first- and second-generation squark masses, while the stops and sbottoms are light. We discuss the implications of this mechanism for the superparticle spectrum and for flavour precision observables. We also provide an explicit realization in a model with simultaneous SUSY and $\SUF$ breaking.
}
\end{center}
\end{titlepage}

\setcounter{page}{2}

\section{Introduction}

In gauge-mediated supersymmetry breaking, the SUSY-breaking hidden sector is charged under the gauge interactions of the supersymmetric Standard Model, and soft terms are induced by gauge boson, gaugino, and hidden-sector loops. This mediation mechanism is attractive because it is predictive and well controlled: The soft terms for the visible sector depend on just a few parameters, and the underlying theory can be a four-dimensional, renormalizable (but typically strongly coupled) quantum field theory. 

The automatic flavour universality of gauge-mediated soft terms is  a major advantage of gauge mediation, since it explains the absence of disastrous squark- and slepton-induced flavour changing neutral currents. It is becoming less attractive in the light of the results from the first LHC run, which point towards first- and second-generation squarks heavier than 0.8 -- 1.8 TeV~\cite{ATLASsqgl, CMSsqgl} for decoupled to equal mass gluinos, respectively. The constraints on third-generation squarks are much weaker by comparison, for example 300 GeV stops are still allowed for LSP masses above 120 GeV~\cite{ATLAS3rd,CMS3rd}. Moreover, light stops are often argued to be preferred by naturalness. A factor of two or more between the squark masses of the first and third generation is clearly at odds with flavour universality, even when taking into account the mass splittings that are induced by renormalization group running from the mediation scale to low energies. Additionally, it has been shown that the 
radiatively induced 
splittings do not ameliorate the fine-tuning problem~\cite{Papucci:2011wy}.

Recently several models have been proposed which allow for flavour non-universal soft masses while retaining most of the predictivity of pure gauge mediation. In \cite{Shadmi:2011hs,Kang:2012ra,Albaid:2012qk, Abdullah:2012tq,Calibbi:2013mka,Galon:2013jba}, messenger fields were allowed to couple to, and mix with, the visible sector matter and Higgs fields in the superpotential. This may give additional non-universal contributions to the scalar soft masses. If the matter-messenger couplings are controlled by suitable flavour symmetries, FCNCs can still be suppressed sufficiently. When the first- and second- generation squarks are split due to the alignment of quark and squark mass matrices~\cite{Galon:2013jba}, this results in significantly weaker limits from direct LHC searches~\cite{Mahbubani:2012qq}. In \cite{Craig:2012yd, Kahn:2013pfa}, an $\SUF$ subgroup of the spurious $\SU{3}^3$ flavour symmetry of the quark sector was gauged and taken to be higgsed by the Yukawa couplings. Its contributions to gauge 
mediation for the various squark masses then depends on the corresponding higgsing scales. For a suitably chosen scale of SUSY breaking mediation, large first- and second-generation squark masses can be induced while keeping the third 
generation light. Similiar models based on abelian flavour symmetries were proposed earlier in \cite{Kaplan:1998jk,Kaplan:1999iq}. 

In the present paper we investigate an alternative possibility to obtain non-universal squark masses from a gauged flavour symmetry. In our model, supersymmetry breaking and flavour breaking are not disconnected, but are triggered by the same vacuum expectation values. This induces tree-level SUSY breaking masses for the broken gauginos, which in turn generate flavour non-universal soft masses through loops. Such ``gauge messenger models'', where massive gauge multiplets couple directly to SUSY breaking, have been considered previously, mainly in the context of GUT breaking (see e.g.~\cite{Dimopoulos:1982gm,Kaplunovsky:1983yx,Murayama:1997pb,Dimopoulos:1997ww,Agashe:1998kg} for early work, and more recently \cite{Dermisek:2006qj,Buican:2009vv,Intriligator:2010be,Matos:2010ie,Bajc:2012sm}). To our knowledge the present model is the first which investigates the effects of gauge messengers for a spontaneously broken gauged flavour symmetry, or in fact for any extension of the Standard Model gauge group by a 
simple factor.

The dominant contribution of gauge messengers to the soft term spectrum is a tachyonic scalar soft mass squared which is generated at one loop. In a model which also contains ordinary chiral messenger fields charged under the SM gauge group, the one-loop tachyon can compete with the usual positive two-loop scalar masses, provided that the $\SUF$ gauge coupling is somewhat smaller than the Standard Model gauge couplings. Since the supersymmetry breaking VEV is aligned with the top and bottom Yukawa couplings in flavour space, the negative contribution to the third-generation squark masses is naturally much larger than the contributions to the first two generation squark masses. This leads to light stop and sbottom squarks. 

This paper is organized as follows: In the next section we precisely define the class of models we are investigating, and present the leading-order effect of flavour gauge messenger fields on the soft terms. In Section \ref{sec:softterms} we discuss the resulting superpartner mass spectra. We illustrate the effect of flavour gauge messengers using a number of parameter points in the MSSM and in the NMSSM. Section \ref{sec:flmod} is concerned with explicit example models for flavour and SUSY breaking: We show that the alignment between flavour symmetry breaking and SUSY breaking, which is a crucial ingredient in our models, can be realized in a simple model. Using this flavour-breaking pattern to generate realistic Yukawa textures, we can then compute the resulting contributions to flavour-changing neutral currents. We summarize our findings and conclude in Section \ref{sec:conclusions}.

\section{Flavour gauge messengers in gauge mediation}\label{sec:flgm}

The matter superfields of the supersymmetric Standard Model transform under an $\SU{3}^5$ non-abelian flavour symmetry when the Yukawa couplings are switched off. Our main interest is an $\SUF$ subgroup under which the quark superfields $Q$, $U$ and $D$ each transform as a ${\bf 3}$.\footnote{Other flavour groups and representations, such as $\SU{3}_{\rm F,L}\times\SU{3}_{\rm F, R}$ with $Q\sim (\bar{\bf 3},{\bf 1})$, $U\sim({\bf 1},{\bf 3})$ and $D\sim({\bf 1},{\bf 3})$, might also be of interest.} We restrict ourselves to the quark sector here, although our construction could easily be extended to a model of lepton flavour, e.g.~in order to embed it into a GUT model. The class of models we are considering is characterized by three essential features:
\begin{itemize}
 \item $\SUF$ is a gauge symmetry,
 \item it is spontaneously broken to $\SU{2}_{\rm F}$ at a scale $M$, where the third-generation Yukawa couplings are generated (while $\SU{2}_{\rm F}$ is broken completely at some lower scale, thus generating the remaining Yukawa couplings),
 \item some of the vacuum expectation values which break $\SUF\into\SU{2}_{\rm F}$ also break supersymmetry.
\end{itemize}
Gauged quark flavour symmetries have been considered in supersymmetric model building for a long time (see e.g.~\cite{Delgado:2012rk, Craig:2012yd,D'Agnolo:2012ie,Krnjaic:2012aj,Cohen:2012rm,Hardy:2013uxa,Huh:2013hga,Kahn:2013pfa,Franceschini:2013ne,Csaki:2013we,Auzzi:2012dv,Bharucha:2013ela,Dudas:2013pja} for some recent work). Among them $\SUF$ is distinguished by being anomaly free with respect to the Standard Model gauge groups, so no new chiral matter with Standard Model charges needs to be added to promote it to a gauge symmetry. The idea of
an approximate $\SU{2}_{\rm F}$ flavour symmetry acting on the first two generations also has a long history \cite{Dine:1993np,Pomarol:1995xc,Barbieri:1995uv,Barbieri:1997tu, Barbieri:2011ci}. What is new here is mainly the third point: The same dynamics that leads to $\SUF\into\SU{2}_{\rm F}$ breaking may also be responsible for supersymmetry breaking. Later we will construct an explicit model where this mechanism is realized. For now we focus on the consequences for the squark soft term spectrum.

When a gauge symmetry such as $\SUF$ is higgsed, the gauge fields and gauginos associated to the broken gauge generators become massive. If the breaking is non-supersymmetric, in the sense that some charged fields acquire $F$-term vacuum expectation values, this will lead to tree-level SUSY breaking mass splittings between the broken gauge fields and gauginos. Thus they become messenger fields for gauge-mediated supersymmetry breaking, inducing soft masses for the fields that are charged under $\SUF$ through loops. (When allowing for nonzero $D$-terms, they can even induce soft masses at the tree level \cite{Nardecchia:2009ew}, but here we will only consider models in which the $D$-terms vanish.)

Gauge messengers for some general gauge group $G$ broken to $H\subset G$ were studied in great detail in \cite{Intriligator:2010be}. This analysis was conducted using a formalism similar to general gauge mediation \cite{Meade:2008wd}, which relies only on the assumption that the theory should be perturbative in the gauge coupling $g$.  The SUSY-breaking hidden sector itself, on the other hand, may be strongly coupled as one might expect for a realistic model of dynamical SUSY breaking. In \cite{Intriligator:2010be} it was established that the leading-order effect in $g$ on the visible-sector soft terms is a one-loop scalar soft mass
\be\label{mPhisq}
m_\Phi^2=g^2\,\T^a\T^b \int \frac{d^4p}{(2\pi)^4}\frac{1}{p^2}\Xi^{(0)ab}(p^2)\,.
\ee
Here $\T^a$ are the generators of $G$ in the representation under which $\Phi$ transforms, and $\Xi^{(0)}$ is the ${\cal O}(g^0)$ piece, taken in a limit where $g$ becomes small but the gauge boson mass is kept constant, in the supertraced gauge supermultiplet propagators
\be
\Xi^{ab}(p^2)=\Delta_0^{ab}(p^2)-4\,\Delta_{1/2}^{ab}(p^2)+3\,\Delta_1^{ab}(p^2)
\ee
where
\be\begin{split}
i\vev{D^a(p) D^b(-p)}=&\,\Delta_0^{ab}(p^2)\,,\\
i\vev{\lambda^a_\alpha(p) \bar\lambda^b_{\dot\alpha}(-p)}=&\,\frac{p_{\alpha\dot\alpha}}{p^2}\,\Delta_{1/2}^{ab}(p^2)\,,\\
i\vev{V_\mu^a(p) V_\nu^b(-p)}=&\,\left(\eta_{\mu\nu}-\frac{p_\mu p_\nu}{p^2}\right)\,\Delta_{1}^{ab}(p^2)\,.
\end{split}
\ee
The precise form of $\Xi^{(0)}$ is model-dependent, and incalculable if the hidden sector is strongly coupled, but not essential for our purposes. However, it is important to note that the integral in Eq.~\eqref{mPhisq} is typically negative. This  has been shown to hold quite generally under certain weak assumptions \cite{Intriligator:2010be}, but is easiest to see explicitly when SUSY breaking is small, i.e.~when the SUSY-breaking mass splittings within the gauge supermultiplet are much smaller than the gauge boson and gaugino masses themselves

Let us consider the small SUSY breaking case, where $G$ is broken to $H$ by the lowest- and highest- (i.e.~$F$-term) component VEVs of some chiral superfields, and the SUSY breaking scale $\sqrt{F}$ is suppressed compared to the mediation scale $M$ as set by the supersymmetric VEVs. Then the massive vector superfields can be integrated out supersymmetrically, and the leading effects of SUSY breaking mediation can be computed using the one-loop effective K\"ahler potential \cite{Grisaru:1996ve} (see also \cite{Giudice:1997ni})
 \be\label{Keff}
K_{\rm eff}^{(1)}=\frac{1}{16\pi^2}\tr\left({\cal M}^2_V\log\frac{{\cal M}^2_V}{\Lambda^2}\right)\,.
\ee
Here the mass matrix ${\cal M}_V$ for a massive vector field $V$ is given by
\be
{\cal M}^2_{Vab}=\left.\frac{\partial^2}{\partial V^a\partial V^b}\sum_I\Phi_{I}^\dag\exp\left(g\,V^c \T^c_I\right)\Phi_{I}\right|_{V=0}
\ee
where $I$ runs over all charged chiral superfields $\Phi_I$, and the $\T^a_I$ are the generators of the corresponding representation, with $a=1,\ldots,\dim G$. Splitting the $\Phi_I$ into visible chiral superfields $Q_I$ (which do not acquire vacuum expectation values) and hidden fields $Z_i$ (which may acquire vacuum expectation values in both their lowest and $F$-components), Eq.~\eqref{Keff} is seen to contain a term
\be\label{softmassKeff}
K_{\rm eff}^{(1)}= \sum_I Q_{I}^\dag \mathbf{T}_I^{ab} Q_{I}\;{\cal Z}^{ab}+\ldots
\ee
where
\be\label{ZT}
 {\cal Z}^{ab}=\frac{g^2}{16\pi^2}\log\left(g^2\sum_i \frac{Z_{i}^\dag \Tb_i Z_{i}}{\Lambda^2}\right)^{ab}\,,\qquad\mathbf{T}^{ab}=\{\T^a,\T^b\}\,.
\ee
The $\theta^2\bar\theta^2$ component of ${\cal Z}^{ab}$ will contribute to the scalar soft masses at one loop,
\be\label{mQIsq}
\delta m_{Q_I}^{2\text{ (1-loop)}}=-\left.\mathbf{T}_I^{ab}{\cal Z}^{ab}\right|_{\theta^2\bar\theta^2}\,.
\ee
As already emphasized, these contributions are generally tachyonic, and non-vanishing if there are several $Z_i$ with non-vanishing VEVs. In the case that all $Z_{i}^\dag \Tb_i Z_{i}$ VEVs commute, this is seen by expanding the logarithm in Eq.~\eqref{ZT} to obtain \cite{Intriligator:2010be}
\be\label{mQIsq2}
\delta m_{Q_I}^{2\text{ (1-loop)}}=-\frac{g^2}{16\pi^2}\Tb_{I}^{ab}\left(\frac{(F^\dag,F)(Z^\dag,Z)-(F^\dag,Z)(Z^\dag,F)}{(Z^\dag,Z)^2}\right)^{ab}+{\cal O}(|F|^4)
\ee
where the inner products are defined in terms of the highest- and lowest-component VEVs $F_{Z_i}$ and $Z_i$ by
\be
(F^\dag,F)^{ab}=\sum_i {F_{Z_i}}^\dag\, \Tb_i^{ab}\, F_{Z_i}\,\qquad (F^\dag,Z)^{ab}=\sum_i {F_{Z_i}}^\dag\,\Tb_i^{ab}\,Z_i\,\qquad (Z^\dag,Z)^{ab}=\sum_i {Z_i}^\dag\, \Tb_i^{ab}\, {Z_i}\,.
\ee

We are interested in the case where $G=\SUF$ is a quark flavour symmetry with gauge coupling $\gF$, and $H=\SU{2}_{\rm F}$ is the subgroup preserved by switching on only the top Yukawa coupling. The simplest way to break 
$\SUF$ with realistic Yukawa matrices is to use two spurions $\Sigma$, $\Sigma'$ in the $\bar{\bf 6}$ of $\SUF$ (see e.g.~\cite{Craig:2012yd}). The quark superpotential is
\be
W=\frac{\Sigma}{\Lambda} H_u QU+\frac{\Sigma'}{\Lambda} H_d QD
\ee
with $\vev{\Sigma}/\Lambda=Y_u$ and $\vev{\Sigma'}/\Lambda=Y_d$. The simplest way to break $\SUF$ with an $F$-term is to use a spurion $X$ in the ${\bf 3}$. If  $\vev X=(0,\,0,\, F_X\theta^2)^T$ in a basis where $\vev{\Sigma}$ is diagonal and $\vev{\Sigma}_{33}/\Lambda=y_t$, then $\vev X$ preserves $\SU{2}_{\rm F}$.  Eqns.~\eqref{ZT} and \eqref{mQIsq} yield
\be
\delta m_{Q_I}^2=-\frac{\gF^2}{16\pi^2}\frac{|F_X|^2}{|\Sigma_{33}|^2}\;\left(\begin{array}{ccc}\frac{13}{24} & 0 & 0 \\ 0 & \frac{13}{24} & 0 \\ 0 & 0 & \frac{7}{6} \end{array}\right)
\ee
for any of the visible-sector fields $Q_I=\{Q,U,D\}$ transforming as ${\bf 3}$ under $\SUF$, up to corrections suppressed by small Yukawa couplings and CKM angles.

This model lacks an explanation for the flavour hierarchies, as well as a dynamical mechanism to align the SUSY-breaking $F$-term with the third generation in flavour space. Our main example will therefore use a different set of spurions, namely, $\{Z_i\}=\{T,\wt T,S,\wt S,X,\wt X\}$, with untilded fields transforming as ${\bf 3}$ and tilded ones as $\bar{\bf 3}$. The dominant VEVs are
\be\begin{split}
\vev{T}&=\left(\begin{array}{c} 0 \\ 0 \\ v \end{array}\right)\,,\quad\vev{X}=\left(\begin{array}{c} 0 \\ 0 \\ F_X\,\theta^2 \end{array}\right)\,,\\
\quad \vev{\wt T}&=\left(\begin{array}{ccc} 0 & 0 & v^* \end{array}\right)\,,\quad \vev{\wt X}=\left(\begin{array}{ccc} 0 & 0 & F_X^*\,\theta^2 \end{array}\right)\,,
\end{split}
\ee
and the top Yukawa coupling is generated by the operator
\be
W=\frac{\wt T\wt T}{\Lambda^2}H_uQU\,.
\ee
The remaining Yukawa couplings are induced by subdominant supersymmetric VEVs for $S$ and $\wt S$, as we will explain in detail in Section \ref{sec:flmod}. In that section we will also offer a dynamical explanation for the alignment of $X$ and $T$. For this model, we find from Eqns.~\eqref{ZT} and \eqref{mQIsq}, again up to small corrections,
\be\label{msq}
\delta m_{Q,U,D}^2=-\frac{\gF^2}{16\pi^2}\frac{|F_X|^2}{|v|^2}\;\left(\begin{array}{ccc}\frac{7}{6} & 0 & 0 \\ 0 & \frac{7}{6} & 0 \\ 0 & 0 & \frac{8}{3} \end{array}\right)\,.
\ee

The relative mass splittings in Eq.~\eqref{msq} have a simple group-theoretic origin \cite{Intriligator:2010be}. Since in this model all spurions transform in the same representation (up to conjugation) and all VEVs are aligned, ${\cal Z}^{ab}$ in Eq.~\eqref{ZT} is universal for all broken generators and can be chosen as  ${\cal Z}^{ab}={\cal Z}\delta^{ab}$, or more generally $\Xi^{(0)ab}=\Xi^{(0)}\delta^{ab}$ in Eq.~\eqref{mPhisq}. Then Eq.~\eqref{mPhisq} becomes
\be\label{mPhisquni}
\delta m_\Phi^2=g^2\,\Delta c_\Phi \int \frac{d^4p}{(2\pi)^4}\frac{1}{p^2}\Xi^{(0)}(p^2)\,,
\ee
with $\Delta c_\Phi$ the difference between the quadratic Casimirs of the $G$-representation and the $H$-representation of $\Phi$.
For $G=\SUF$, $H=\SU{2}_{\rm F}$, and $g=\gF$ the $\SUF$ gauge coupling, we have
\be\begin{split}
\Delta c_\Phi=&\,\frac{4}{3}-\frac{3}{4}=\frac{7}{12}\,,\qquad \text{first- and second-generation squarks}\,,\\
\Delta c_\Phi=&\,\frac{4}{3}-0=\frac{4}{3}\,,\qquad \text{third-generation squarks}\,.
\end{split}
\ee
Therefore, the one-loop contribution to the squark mass-squared matrices can be written as
\be\label{deltamsq}
\delta m^2_{Q,U,D}=-\frac{\gF^2}{16\pi^2}\,\LamF^2\,\left(\begin{array}{ccc}\frac{7}{6} & 0 & 0 \\ 0 & \frac{7}{6} & 0 \\ 0 & 0 & \frac{8}{3}\end{array}\right)\,,
\ee
where $\LamF$ is some model-dependent characteristic mass scale; in the small SUSY breaking case, $\LamF=|F_X|^2/|v|^2$ and we recover Eq.~\eqref{msq}.

Eq.~\eqref{deltamsq} must be interpreted with some care. First, it holds only at the scale of $\SUF$ breaking, and second, it holds only in a particular flavour basis. Rotating to the super-CKM basis will induce corrections, including small off-diagonal squark masses, which depend on the details of flavour symmetry breaking.

There are other soft terms induced by gauge messenger fields, but these will generically appear only at higher order in perturbation theory. For instance, gauge messengers induce one-loop $A$-terms, but $A\sim \gF^2\,\LamF/(16\pi^2)$ is evidently subdominant with respect to the one-loop soft mass $m_\Phi\sim \gF\,\LamF/(4\pi)$. There are also additional two-loop contributions to the scalar soft masses, and MSSM gaugino masses generated at three-loop order. For the rest of this paper, we will neglect these higher-order effects,\footnote{They may be relevant in models where the one-loop soft mass squared of Eq.~\eqref{mPhisquni} is suppressed for some reason. This is the case when the VEV of the scalar superpartner of the Goldstino is the only \cite{Giudice:1997ni} or more generally the dominant \cite{Intriligator:2010be} source of $\SUF$ breaking. We will not consider such models here.} and retain only the one-loop soft mass of Eq.~\eqref{deltamsq}. Indeed we will eventually take the $\SUF$ gauge coupling to 
be very small, $\gF\approx $ few $\times 10^{-2}$, in order to obtain a realistic phenomenology, so higher loop orders can be safely neglected.

\section{Soft terms and low-energy spectrum}\label{sec:softterms}

Clearly, the soft parameters induced by gauge messengers alone cannot account for a realistic superpartner mass spectrum: The squarks are tachyonic, and gaugino, slepton, and Higgs masses are tiny because they are induced only at higher loop order. We therefore need to consider more general models of gauge mediation where there are also contributions to the soft masses from hidden-sector states charged under $\SU{3}_{\rm C}\times\SU{2}_{\rm L}\times\U{1}_{\rm Y}$. The simplest such models are models with weakly coupled chiral messenger superfields, such as minimal gauge mediation. For concreteness, let us therefore assume that the matter and gaugino soft masses are as predicted by minimal gauge mediation (see \cite{Giudice:1998bp} for a review) at the messenger scale $M$, i.e.~given in terms of an $\SU{5}$ messenger index $N_5$ and the scale $|\Lambda_{\rm MGM}|\sim |F/M|$ (taken to satisfy $|\Lambda_{\rm MGM}|\ll M$) as
\be\begin{split}\label{mgmmasses}
M_{1,2,3}=&\;\frac{g_{1,2,3}^2}{16\pi^2}\,\Lambda_{\rm MGM}\,N_5\,,\\
m_{Q}^2=&\;\left(\frac{1}{16\pi^2}\right)^2\left[\frac{8}{3}g_3^4+\frac{3}{2}g_2^4+\frac{1}{30} g_1^4\right]\,\left|\Lambda_{\rm MGM}\right|^2\,N_5\,\mathbbm{1}\,,\\
 m_{U}^2=&\;\left(\frac{1}{16\pi^2}\right)^2\left[\frac{8}{3}g_3^4+\frac{8}{15}g_1^4\right]\left|\Lambda_{\rm MGM}\right|^2\,N_5\,\mathbbm{1}\,,\quad
m_{D}^2=\left(\frac{1}{16\pi^2}\right)^2\left[\frac{8}{3}g_3^4+\frac{2}{15}g_1^4\right]\left|\Lambda_{\rm MGM}^2\right|\,N_5\,\mathbbm{1}\,,\\
m_{L}^2=&\;\left(\frac{1}{16\pi^2}\right)^2\left[\frac{3}{2} g_2^4+\frac{3}{10} g_1^4\right]\left|\Lambda_{\rm MGM}\right|^2\,N_5\,\mathbbm{1}\,,\quad
m_{E}^2=\left(\frac{1}{16\pi^2}\right)^2\left[\frac{6}{5} g_1^4\right]\left|\Lambda_{\rm MGM}\right|^2\,N_5\,\mathbbm{1}\,.\\
\end{split}
\ee
To these we add the gauge messenger contributions to the squark masses of Eq.~{\eqref{deltamsq} 
\be\label{gmmsoft}
\delta m_{Q,U,D}^2=-\frac{\gF^2}{16\pi^2}\,\LamF^2\,\left(\begin{array}{ccc}\frac{7}{6} & 0 & 0 \\ 0 & \frac{7}{6} & 0 \\ 0 & 0 &\frac{8}{3}\end{array}\right)\,.
\ee
We emphasize however that our mechanism as such does not rely on minimal gauge mediation: Similar conclusions will be reached whenever one assumes that the squark masses are flavour-blind (as they generally are in conventional gauge mediation without gauged flavour symmetries) except for the gauge messenger contributions of Eq.~\eqref{gmmsoft}. In particular, Eqs.~\eqref{mgmmasses} could be replaced by the soft masses obtained from any model of general gauge mediation. Moreover, the mediation scales for the chiral and gauge messengers could in general be distinct.

\begin{figure}
 \centering
 \includegraphics[width=.6\textwidth]{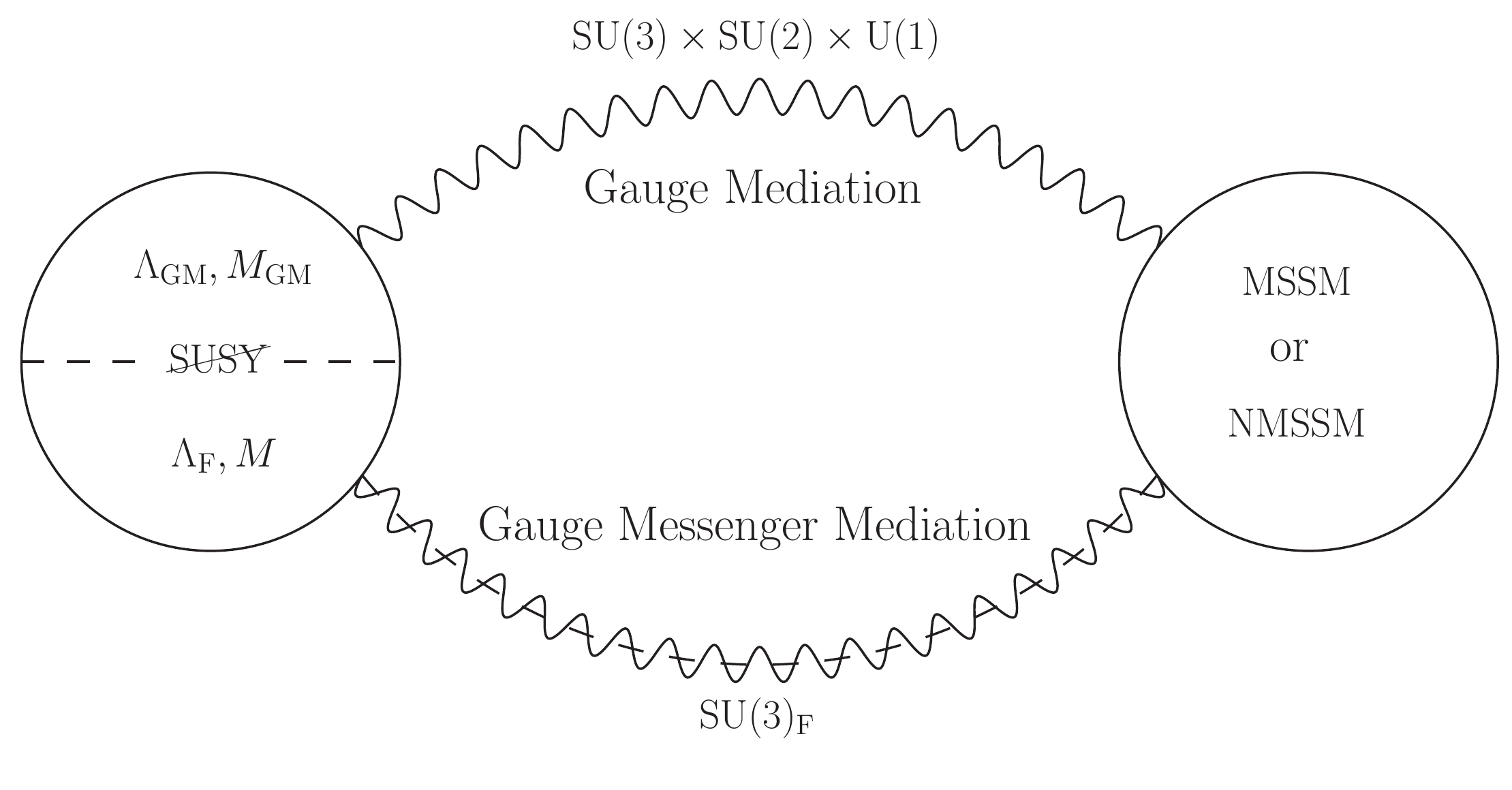}
  \caption{A sketch of the model we are analysing. SUSY breaking is mediated to the visible sector both by Standard Model gauge interactions (for instance, via ordinary chiral messenger superfields) and by the higgsed $\SUF$ (via its massive vector superfields).}
\end{figure}

Assuming that $\LamF$ is comparable with $\Lambda_{\rm MGM}$, the effect on the spectrum will mostly depend on the size of the extra gauge coupling $\gF$. If $\gF$ is of the order of the Standard Model gauge couplings or larger, the tachyonic one-loop squark masses of Eq.~\eqref{gmmsoft} will be dominant over the positive two-loop squark masses of Eqs.~\eqref{mgmmasses}, leading to an unrealistic spectrum. On the other hand, if $\gF$ is too small, there will be no noticeable effect coming from the gauge messengers at all. The most interesting parameter region is the one where the stop and sbottom masses from Eqs.~\eqref{gmmsoft} and \eqref{mgmmasses} are of similar magnitude. This is typically the case for $\gF\approx$ few $\times 10^{-2}$, whereupon the stop and sbottom squarks become light, while the first and second generation squarks are less affected.

A well-known benefit of large stop masses is of course that they allow one to accommodate a 125 GeV Higgs boson within the MSSM. This is because the lightest Higgs mass receives loop corrections proportional to $\log( m_{\tilde t_1} m_{\tilde t_2}/m_t^2)$. Another potentially large correction comes from the stop trilinear parameter $A_t$. However, it is well known to be difficult to obtain a 125 GeV Higgs within pure gauge mediation, because $A_t$ is predicted to be negligibly small at the mediation scale. Lifting the lightest Higgs mass with only the radiatively induced $A_t$ then requires extremely heavy $m_{\tilde t}$. These observations would thus seem to disfavour our gauge messenger model in connection with the MSSM. 

It is important to note that is in fact not the case, since these arguments rest on rather too strong assumptions about SUSY breaking mediation. Within potentially realistic scenarios, our gauge messenger contribution to the stop mass may indeed make it easier to obtain a 125 GeV Higgs without 
having to resort to extreme parameter values. The crucial point here has actually been known for some time, although it is often ignored (as evidenced by the fact that phenomenological studies of ``GMSB'' benchmark scenarios are still being conducted): \emph{Pure gauge mediation has a $\mu/B\mu$ problem \cite{Dvali:1996cu}; a mechanism which solves this problem will generically give additional contributions to the Higgs soft masses and trilinear terms on top of the purely gauge-mediated ones.} Here by pure gauge mediation we mean any model in which the visible and hidden sector are coupled only by Standard Model gauge interactions. Then the higgsino mass parameter $\mu$ vanishes, as does the Higgs mass mixing parameter $B\mu$ at the messenger scale.\footnote{There is a way to avoid this conclusion if one assumes that the origin of $\mu$ is unrelated to supersymmetry breaking, that it happens to be of the order of the soft mass scale by accident, and that $B\mu$ at lower scales is induced radiatively. 
We will 
not consider 
this possibility as it 
leaves an unnatural 
coincidence of scales unexplained.} To obtain realistic $\mu$ and $B\mu$ terms, additional interactions between the Higgs sector and the SUSY-breaking hidden sector are needed, but these will affect also $m_{H_u}^2$, $m_{H_d}^2$ and the trilinear terms in a model-dependent manner. For phenomenological studies of gauge mediation, it is therefore preferable to either rely on an explicit model which realizes this (and which ideally should allow one to calculate the resulting soft terms), or to leave all Higgs sector soft terms as free parameters. 

It is highly nontrivial to build a calculable model which solves the $\mu$-$B\mu$ problem in gauge mediation, and the Higgs sector is not actually the focus of our study. We therefore choose to treat $\mu$, $B\mu$, $m_{H_u}^2$, $m_{H_d}^2$, and $A_t$ as independent parameters, with the understanding that they could emerge from a variation of any of the more complete models on the market (see e.g.~\cite{Giudice:2007ca,Craig:2012xp}). By contrast, the soft terms in the matter and gaugino sectors are taken as predicted by minimal gauge mediation with additional $\SUF$ gauge messengers.

To match to the Standard Model at low energies, the model parameters must be chosen such that both the electroweak scale and the lightest Higgs mass $m_{h^0}=125$ GeV are reproduced properly. 
In addition, the soft terms should be chosen such as not to be in conflict with LEP and LHC search bounds. This places severe constraints on the spectrum, in particular on the masses of the first two generation squarks and of the gluino, all of which should be significantly above a TeV.

Naturalness arguments, on the other hand, favour stop and gluino masses which are as low as possible. In the MSSM, the most natural remaining parameter region is characterized by sub-TeV stop squarks, with the Higgs mass accounted for by a maximal contribution from stop mixing. This in turn requires $|A_t|\approx 2\,M_S$ (where $M_S^2\equiv m_{\tilde t_1} m_{\tilde t_2}$). As we have argued above, a realistic gauge-mediated model supplemented with additional Higgs-hidden sector interactions may well allow for large $A$-terms. Usually, however, it does not allow for reasonably light stops while at the same time evading the LHC bounds on the first generation squarks and gluinos. This is where the $\SUF$ gauge messenger contributions can play a crucial role.

\begin{figure}
 \centering
 \includegraphics[width=.48\textwidth]{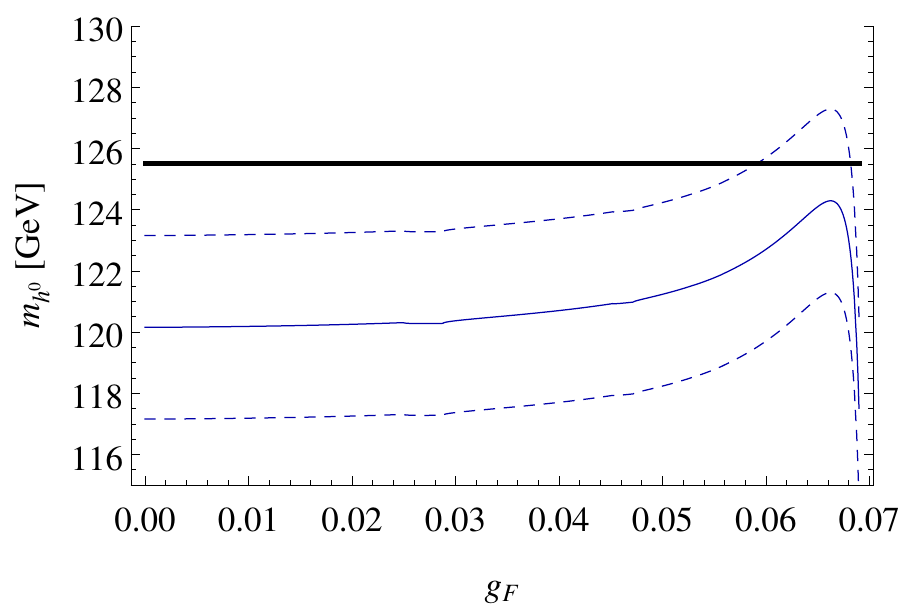}$\quad$
 \includegraphics[width=.48\textwidth]{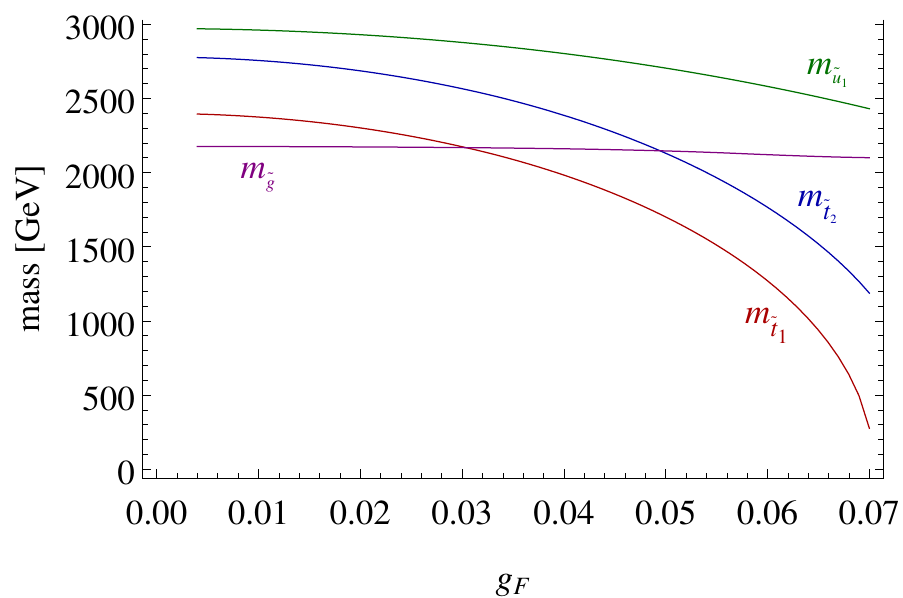}
  \caption{Lightest Higgs mass (left) and various soft masses (right) as a function of $\gF$ for a sample parameter point, as computed with {\tt SOFTSUSY 3.3.7} \cite{Allanach:2001kg} and {\tt
FeynHiggs 2.9.4} \cite{Heinemeyer:1998yj, Degrassi:2002fi,Frank:2006yh}. The model parameters are listed in Table \ref{benchmarks} under MSSM-I.  The blue dotted lines in the left panel show the $\pm 3$ GeV theory uncertainty interval around the theory prediction, while the solid horizontal line shows the LHC central value of 125.5 GeV. As evident from the right panel, the gauge messenger contribution to the soft masses may significantly affect the third-generation squark masses, allowing for maximal stop mixing with only moderately large $|A_t|$, while the first-generation squarks and the gluino remain heavy enough to evade the LHC bounds.  }\label{mh0msq}
\end{figure}

Fig.~\ref{mh0msq} shows the effect on the squark sector mass parameters in the MSSM, and the consequences for the lightest Higgs mass, as the gauge messenger contributions are switched on. The Higgs sector parameters were chosen to allow for maximal stop mixing when the gauge messenger contribution to the stop mass is sizeable. They are listed under ``MSSM-I'' in Table \ref{benchmarks}. The resulting Higgs mass can be compatible with the LHC discovery when taking theory uncertainties into account.  Of course maximal stop mixing is also possible with no gauge messenger contributions at all, but this would require either extremely large $A$-terms (of the order of $5$ TeV for the parameter point we are showing) or dangerously small first-generation squark and gluino masses (since they are tied to the stop masses in gauge-mediated models without gauge messengers).

\begin{table}[ht]
\centering
\begin{tabular}{cccc}
&  MSSM-I & MSSM-II & NMSSM \\[3pt] \hline
$\Lambda_{\rm F}, \Lambda_{\rm MGM}$ & $3\times 10^5$ GeV & $10^5$ GeV & $2.2\times 10^5$ GeV \\
$M$ & $10^7$ GeV & $10^{12}$ GeV & $10^7$ GeV \\
$N_5$ & $1$ & $3$ & $1$ \\
$A_0$ & $-2000$ GeV & $0$ & $0$ \\
$m_{H_u}^2$ & $10^5\,($GeV$)^2$ & $-1.8\times 10^6\,($GeV$)^2$ & $(10^4\to 10^6)\,($GeV$)^2$ \\
$m_{H_d}^2$ & $10^5\,($GeV$)^2$ & $10^5\,($GeV$)^2$ & $(10^4\to 10^6)\,($GeV$)^2$ \\
$\left.\tan\beta\right|_{m_Z}$ & $10$ & $10$ & $(1\to 5)$ \\
$v_S$ &&& $(400\to 2000)\,$ GeV\\
$\kappa$ &&& $(0.55\to 1)$\\
$\lambda$ &&& $(0.55\to 1)$\\
$A_\kappa, A_\lambda$ &&& 0
\end{tabular}
\caption{Model parameters at the mediation scale $M$ for the three cases we are discussing: The MSSM with a large $A$-term, which could be induced by Higgs-messenger superpotential couplings (MSSM-I); the MSSM with a radiatively induced large $A$-term, necessiating a heavy gluino and a high mediation scale (MSSM-II); the NMSSM, where we scan over the Higgs sector parameters in a suitable range. For completeness, we mention also the off-diagonal squark mass term Eq.~\eqref{deltamsq23}, for which we chose $\eta=1$ and $\epsilon=0.1$; however this has a negligible effect on the spectrum and will only be important later on when we discuss flavour violation.}\label{benchmarks}
\end{table}

In the left panel of Fig.~\ref{RG} we show the RG evolution of the stop and Higgs sector soft masses from the mediation scale to the TeV scale, for the same parameter point but keeping $\gF=1/15\approx g'/5$ fixed. Note that the lighter stop soft mass, roughly given by $m_{U\,33}^2$, is negative at high energies (this is also the case for $m_{D\,33}^2$; all other squark masses are positive at all scales). When running down towards the electroweak scale, it is driven positive by the gluino mass. Tachyonic boundary conditions for the stops have previously been employed to improve the fine-tuning in the MSSM \cite{Dermisek:2006ey}, in particular also in the context of gauge mediation \cite{Draper:2011aa} and an $\SU{5}$-based gauge messenger model \cite{Dermisek:2006qj}. 

For a generic direction in the space of the MSSM scalar fields, a negative running soft mass $\sim - m_{\rm soft}^2$ at high renormalization scales $Q\gg m_{\rm soft}$ is no cause for concern (here $m_{\rm soft}\sim$ 1 TeV denotes the soft mass scale). At first glance it would seem to induce a VEV of the order $v\sim m_{\rm soft}/g^2$, where $g^2$ is some combination of MSSM gauge couplings. However, $v\ll Q$ implies that the running tree-level potential at the scale $Q$ is a poor approximation to the full effective potential, since the higher loop corrections would involve large logarithms. Instead one should use the tree-level potential at the scale $v$, but there all squark masses are positive, so the additional vacuum is in fact spurious.

A potentially problematic case are the $D$-flat directions along which the quartic coupling vanishes, such that a large field expectation value $v \sim Q$ could easily develop  \cite{Falk:1995cq,Ellis:2008mc}. If a mass along these directions becomes negative at large $Q$, the potential would be unbounded from below. In the presence of suitable higher-dimensional operators all $D$-flat directions are lifted \cite{Gherghetta:1995dv}, and the runaway is stabilized, but additional vacua will appear in which electric charge and/or colour are broken. For the above model the most dangerous $D$-flat direction is the one associated with the operator $\tilde t_R\tilde b_R\tilde s_R$, because it involves the two tachyonic fields $\tilde t_R$ and $\tilde b_R$ and only one positive-mass field $\tilde s_R$. We have checked that the mass along this direction remains positive at all scales up to $M$, for all values of $\gF$ that yield a tachyon-free spectrum at the electroweak scale.

\begin{figure}
 \centering
 \includegraphics[width=.48\textwidth]{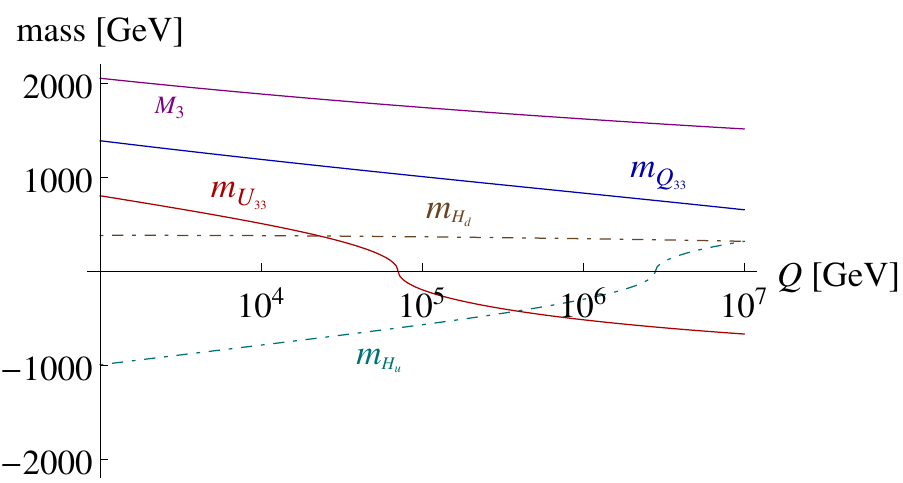}$\quad$
 \includegraphics[width=.48\textwidth]{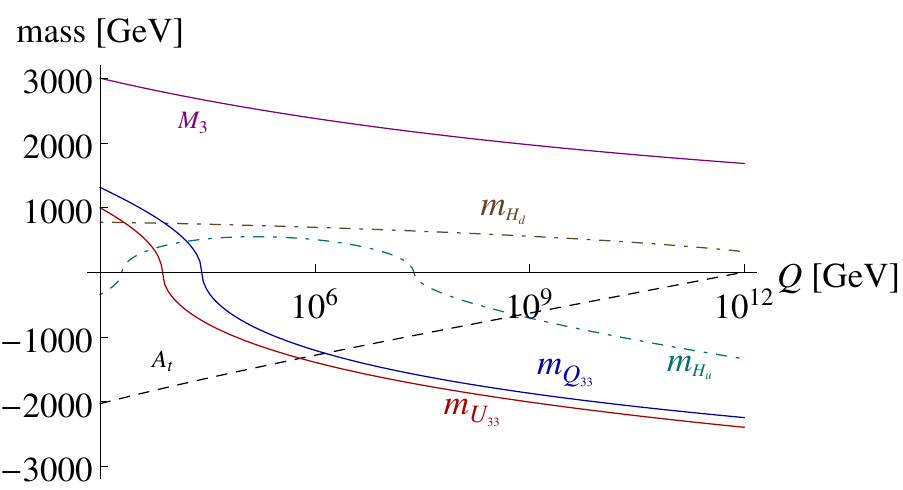}
  \caption{Running Higgs and stop mass parameters for fixed $\gF$, as a function of the renormalization scale $Q$, according to {\tt SOFTSUSY}. (More precisely, what is plotted is sign$(m^2)\sqrt{|m^2|}$.) Left panel: Point MSSM-I of Table \ref{benchmarks} with $\gF=1/15$. Right panel: Point MSSM-II of Table \ref{benchmarks}, with $\gF=3/20$, leading to radiatively induced maximal stop mixing.}\label{RG}
\end{figure}

A somewhat more extreme case is shown in the right panel of Fig.~\ref{RG}, corresponding to the parameters listed under ``MSSM-II'' in Table \ref{benchmarks}. This point serves to show that maximal stop mixing can even be purely radiatively induced in our model, although this comes at the price of a high mediation scale, a rather large (around $3$ TeV) gluino mass, and squarks which become tachyonic starting from around only $10^{4}$ GeV. Radiative effects, in particular due to the gluino mass, eventually drive the squark masses positive and the $A$-term large. Similar soft mass patterns have been discussed in \cite{Draper:2011aa}. For this model, the potential is indeed unbounded from below, which signals the appearance of additional charge- and colour-breaking vacua. These can be problematic in two ways: Firstly, the universe could prefer to settle in them, rather than in the electroweak vacuum, during the early cosmological evolution. Secondly, even if our vacuum is the preferred one, one still needs to 
ensure that it does not decay on cosmological timescales. A detailed investigation of the constraints on negative squark masses from cosmology is beyond the scope of this paper, but would certainly be interesting to conduct (see also \cite{Ellis:2008mc, Reece:2012gi, Camargo-Molina:2013sta}).

Flavour gauge messengers may also be included in extensions of the MSSM where there is no need to rely on large corrections to the Higgs mass from the stop sector. For example, in the NMSSM, a SM-like Higgs with the proper mass can be obtained even with low stop masses and mixings, because there is an additional contribution to the Higgs quartic coupling coming from a superpotential term $\lambda\,SH_u H_d$ with $S$  a gauge singlet. Fig.~\ref{NMSSM} shows the squark masses in a random scan over the parameter space of the NMSSM Higgs sector (see also Table \ref{benchmarks}). For obvious reasons, the dependence of the squark masses on $\gF$ is similar as in the MSSM (Fig.~\ref{mh0msq}); the difference between these plots is, however, that all of the points shown in Fig.~\ref{NMSSM} are compatible with a lightest Higgs mass of $125.5$ GeV.

\begin{figure}
 \centering
 \includegraphics[width=.48\textwidth]{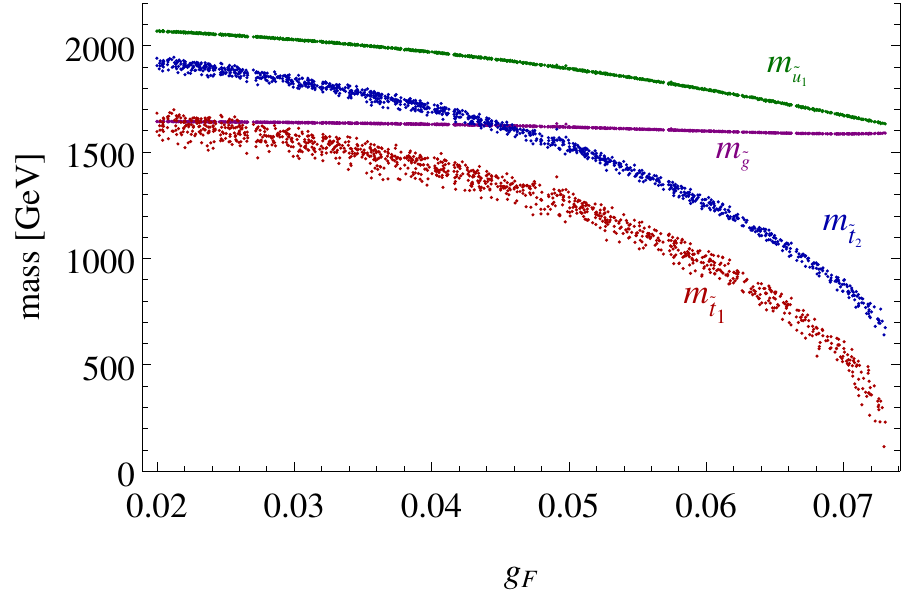}
  \caption{Squark and gluino masses in a random scan over the NMSSM parameter space using {\tt SPheno 3.2.3} \cite{Porod:2003um,Porod:2011nf} and {\tt SARAH 3.3.0} \cite{Staub:2012pb}. The parameters and scan ranges are as in Table \ref{benchmarks} (rightmost column); all points displayed have $m_{h^0}=125.5\pm 3$ GeV}\label{NMSSM}
\end{figure}

Our examples show that it is possible to obtain a gauge-mediated soft term spectrum with light third-generation squarks from flavour gauge messengers, in a variety of scenarios. If hints of supersymmetry were to surface in stop or sbottom searches, this would be a natural way to explain the lightness of the third generation within gauge mediation.

Light stop squarks are often argued to alleviate the supersymmetric little hierarchy problem. This is because the Higgs potential is very sensitive to the stop masses, so if the stop masses are much larger than the electroweak scale, accidental cancellations are required in order to obtain the proper Fermi scale. Conversely, in a model with relatively light stops the electroweak scale can be naturally of the right order. Our mechanism provides an example of how this argument may fail (but fail in interesting way): While we can easily obtain sub-TeV stops, by playing off the positive contribution to the soft mass from standard gauge mediation against the negative contribution from flavour gauge messengers, these two contributions are individually large and independent. The usual measure of fine-tuning is the sensitivity of the electroweak scale with respect to variations of the independent fundamental model parameters. In our case, the electroweak scale depends sensitively on both (large) contributions to the 
stop mass, regardless of whether or not their sum is small, so by this standard we do not gain much in terms of fine-tuning from having light stops.\footnote{Any discussion on the subject of naturalness and fine-tuning, however, relies on assumptions about the UV completion. In~ \cite{Dermisek:2006ey, Dermisek:2006qj} the authors argue that light stops, or tachyonic high-scale boundary conditions for the stop masses, could even be beneficial for naturalness.} Their only benefit regarding naturalness is that, within the MSSM, less extreme values for the $A$-terms are needed to lift the Higgs mass.

\section{An explicit model}\label{sec:flmod}

\subsection{Supersymmetry breaking and flavour symmetry breaking}

The mechanism we have proposed relies on the alignment of supersymmetry breaking and $\SUF\into\SU{2}_{\rm F}$ breaking in flavour space. To show that this can be easily realized, let us construct a simple O'Raifeartaigh model as an example. The superpotential is
\be
W=\kappa\,Y\left(\wt T T-f^2\right)+m\,\wt T X+cm\,\wt X T\,,
\ee 
where $T$ and $X$ are chiral superfields transforming as ${\bf 3}$ under $\SUF$, $\wt T$ and $\wt X$ transform as $\bar {\bf 3}$, and $Y$ is a singlet. There is a $\U{1}_R$ symmetry under which $X$, $\wt X$ and $Y$ carry charge $2$. For later reference we note that there is also a non-$R$ $\U{1}$ symmetry acting on $T$, $\wt T$, $X$ and $\wt X$, with a $\mathbb{Z}_2$ subgroup which will be of interest for us. All fields except $Y$ are odd under this $\mathbb{Z}_2$.

We choose the parameters $\kappa$, $c$, $m$ and $f$ to be real and positive, and such that $\kappa f>m$ and $\kappa f >cm$. Then the $F$-term potential is minimized at
\be\begin{split}\label{TX}
T&\,=(v_1,v_2,v_3)^T\qquad\text{subject to } v_i^* v_i=\left(\frac{f^2}{c}-\frac{m^2}{\kappa^2}\right)\,,\\
\wt T&\,= c\,(v_1^*,v_2^*,v_3^*)\,,\\
X&\,=-\frac{\kappa\,Y}{m}T\,,\\
\wt X&\,=-\frac{\kappa\,\,Y}{cm} \wt T=X^\dag\,,
\end{split}
\ee
with $Y$ a flat direction at tree level. Supersymmetry is broken because
\be\label{FXFY}
\frac{\partial W}{\partial X}=m\,\wt T\neq 0\,,\quad \frac{\partial W}{\partial\wt X}=cm\,T\neq 0\,,\quad \frac{\partial W}{\partial Y}=\kappa\,\left(\wt TT-f^2\right)=-\frac{cm^2}{\kappa}\neq 0\,.
\ee
 The one-loop effective potential will stabilize the remaining tree-level flat directions, with the $Y$ VEV at or close to zero  if the $\SUF$ gauge coupling is small \cite{Witten:1981kv, Intriligator:2007py}. 

With these VEVs, the $\SUF$ $D$-term potential vanishes for $c=1$. For $c\neq 1$ there will be a non-vanishing $D$-term induced by $T$ and $\wt T$. Explicitly, in a gauge where $T=(0,\,0,\,v)^T$, 
\be
D=\left(0,\,0,\,0,\,0,\,0,\,0,\,0,\,\frac{1-c^2}{\sqrt{3}}\;|v|^2\right)\,.
\ee
In the absence of other fields taking VEVs, this $D$-term will push the vacuum away from the $F$-term pseudomoduli space of Eqs.~\eqref{TX}. It is then easy to see that also in the new vacuum the $D$-term will be non-zero, which could induce dangerous VEVs for the squarks. We therefore assume that the overall $D$-term vanishes\footnote{This is a notable difference to models of tree-level gauge mediation \cite{Nardecchia:2009ew}, where a gauge symmetry is also broken by an $F$-term, but the ensuing $D$-term plays a crucial role for generating soft masses.} due to another hidden sector field taking a VEV in the $T$-direction. For instance, if $c<1$, an additional field $\wt Z$ in the $\bar{\bf 3}$ will take a VEV $\wt Z=(0,0,\sqrt{1-c^2}\;v)$, cancelling the $D$-term.

The $F$-terms of $X$ and $\wt X$ break the flavour symmetry, and they are dynamically aligned in flavour space with the VEVs of $T$ and $\wt T$ by the equations of motion. The gauge-mediated soft terms are calculated as outlined in Section \ref{sec:flgm} (see also Appendix A). The result is Eq.~\eqref{deltamsq} with 
\be\label{LambdaF}
\LamF^2=\frac{|F_X|^2 + |F_{\wt X}|^2}{|T|^2+|\wt T|^2+|\wt Z|^2}=c^2m^2\,.
\ee
Here we have neglected the $Y$, $X$, and $\wt X$ VEVs, as they will be small for $\gF\ll\kappa$.

In more general models, especially in strongly coupled ones, the small SUSY breaking limit need not be realized, and the flavour-breaking $F$-terms may be of the same order as the largest VEVs. This case of a single scale for SUSY breaking and gauge symmetry breaking is investigated in \cite{Buican:2009vv} for a $\U{1}$ symmetry instead of $\SUF$, and also in general in \cite{Intriligator:2010be}. While the conclusions of Section \ref{sec:softterms} would remain unaffected, the relation between the VEVs and the scale $\LamF$ would become more complicated than Eq.~\eqref{LambdaF} (which holds only at the leading order in $F_X/T^2$, or equivalently in $m/f$). For our purposes it is sufficient to consider the simpler case of Eq.~\eqref{LambdaF}.

In order to break $\SU{2}_{\rm F}$ completely, and to thereby generate realistic Yukawa matrices, additional fields charged under $\SUF$ should take VEVs which are not aligned with $\vev T$. The simplest possibility is to add another pair of chiral superfields $S$, $\wt S$ in the ${\bf 3}\oplus\bar{\bf 3}$ whose VEVs are generated independently of SUSY breaking (and parametrically smaller than $\vev T$). Then the flavour-breaking $F$-term of $X$ remains aligned with $T$. Superpotential couplings $S\wt X$ or $\wt S X$ would spoil this alignment, but they are forbidden if $S$ and $\wt S$ are even under the $\mathbb{Z}_2$ symmetry. Furthermore, the condition that no $D$-term should arise from the $S$ and $\wt S$ VEVs fixes $\vev{S}$ and $\vev{\wt S^\dag}$ to be equal up to a phase. Eq.~\eqref{deltamsq} will receive small corrections from $\SU{2}_{\rm F}$ breaking; the dominant contribution is calculated in Appendix A.

To also obtain additional soft masses from minimal gauge mediation, the simplest possibility is to add flavour-singlet messenger fields $\Phi$, $\wt\Phi$ which transform as ${\bf 5}\oplus\bar{\bf 5}$ under $\SU{5}\supset\SU{3}_{\rm C}\times\SU{2}_{\rm L}\times\U{1}_{\rm Y}$ and which couple to $Y$ as
\be\label{Wmess}
W=Y\Phi\wt\Phi+M\Phi\wt\Phi\,,
\ee
where $M$ is an explicit messenger mass.\footnote{Note that the $R$-symmetry breaking superpotential of Eq.~\eqref{Wmess} will destabilize the SUSY-breaking vacuum. This is a common problem when trying to extend O'Raifeartaigh models into full models of minimal gauge mediation. If the explicit $R$-breaking is small, the SUSY-breaking minimum may persist as a metastable state, and the model may still be realistic. However, ultimately our SUSY breaking model should be regarded as a stepping stone towards a full model and is meant to merely illustrate the dynamical alignment of the $F$-term with the flavour-breaking VEV.}
Additionally, a number of Standard Model singlets charged under $\SUF$ should be added to cancel the $\SUF^3$ anomaly, and given a mass by coupling them to the $\SUF$-breaking VEVs. Finally, there should also be heavy fields charged under $\SUF$ that are integrated out at a somewhat higher scale, thereby generating the operators which ultimately induce the Yukawa couplings (see next section). We do not specify all these additional states because they will not have any significant effect on the visible sector -- they affect the scalar soft masses only at the two-loop level in $\gF$. For completeness, the field content as far as we have specified it is listed in Table \ref{fields}.

\begin{table}
\centering
 \begin{tabular}{c|clccc}
  field & $\SUF$ & $G_{\rm SM}$ & $\mathbb{Z}_2$ &$\mathbb{Z}_4$ & $\U{1}_R$\\\hline 
$Q$ & ${\bf 3} $ & $({\bf 3},{\bf 2})_{1/6}$ & $+$ & $0$ & $2/3$\\
$U$ & ${\bf 3}$ & $(\bar{\bf 3},{\bf 1})_{-2/3}$ & $+$ & $0$& $2/3$\\
$D$ & ${\bf 3}$ & $(\bar{\bf 3},{\bf 1})_{1/3}$ & $+$ & $0$& $2/3$\\
$H_u$ & ${\bf 1}$ & $({\bf 1},{\bf 2})_{1/2}$ & $+$ & $0$& $2/3$\\
$H_d$ & ${\bf 1}$ & $({\bf 1},{\bf 2})_{-1/2}$ & $+$ & $0$& $2/3$\\
$T$ & ${\bf 3}$ & ${\bf 1}$ & $-$ & $0$& $0$\\
$\wt T$ & $\bar{\bf 3}$ & ${\bf 1}$ & $-$ & $0$& $0$\\
$X$ & ${\bf 3}$ & ${\bf 1}$ & $-$ & $0$ & $2$\\
$\wt X$ & $\bar{\bf 3}$ & ${\bf 1}$ & $-$ & $0$ & $2$\\
$S$ & ${\bf 3}$ & ${\bf 1}$ & $+$ & $1$& $0$\\
$\wt S$ & $\bar{\bf 3}$ & ${\bf 1}$ & $+$ & $2$& $0$\\
$Y$ & ${\bf 1}$ & ${\bf 1}$ & $+$ & $0$ & $2$\\
$\wt Z$ & $\bar{\bf 3}$ & ${\bf 1}$ & &&\\
$\Phi$ & ${\bf 1}$ & $(\bar{\bf 3},{\bf 1})_{1/3}\oplus ({\bf 1},{\bf 2})_{-1/2}$ & $+$ & $0$& $1$\\
$\wt\Phi$ & ${\bf 1}$ & $({\bf 3},{\bf 1})_{-1/3}\oplus ({\bf 1},{\bf 2})_{1/2}$ & $+$  & $0$ & $1$\\
 \end{tabular}
\caption{Field content and representations under $\SUF$, under the Standard Model gauge group, under $\mathbb{Z}_2$, under the $\mathbb{Z}_4$ imposed in Section \ref{sec:yukckm}, and under an approximate $R$-symmetry. The $\wt Z$ charges are largely arbitrary, so long as they are chosen to forbid superpotential couplings between $\wt Z$ and the other fields. Cancelling the $(\SUF)^3$ anomaly requires additional fields which we have not specified.}\label{fields}
\end{table}

We do not advocate this model as a fully realistic hidden sector (for instance, in a full model one would expect all scales to be generated dynamically). However, it does exhibit all the characteristics which we used in Sect.~\ref{sec:softterms}, and these might well be found also in a more complete dynamical model of SUSY and flavour breaking:
\begin{itemize}
 \item There is a step-wise breaking of the flavour symmetry, $\SUF\into\SU{2}_{\rm F}\into 0$, with the two steps triggered by the VEVs of $T$, $\wt T$ and $S$, $\wt S$. 
 \item The SUSY-breaking fields $X$, $\wt X$ take part in the first breaking step. Their lowest components do not develop a significant VEV, but their $F$-terms are aligned with the VEVs of $T$ and $\wt T$ by the equations of motion.
 \item The $F$-terms of $X$ and $\wt X$ induce SUSY-breaking gaugino masses for the broken flavour gauge bosons, which become gauge messengers.
 \item Additional chiral messengers $\Phi,\wt\Phi$ with Standard Model gauge charges also contribute to the visible-sector soft masses.
\end{itemize}

\subsection{Yukawa and CKM hierarchies}\label{sec:yukckm}

The gauge messenger contribution to the squark soft masses will generally induce flavour changing neutral currents, which are strongly constrained by experiment. To calculate the effects of flavour violation, we need to specify the  $\SUF$ breaking pattern in more detail. 

For the sake of concreteness, we will study a simple non-abelian Froggatt-Nielsen-like flavour model as an example.\footnote{For other models also based on an $\SUF$ horizontal symmetry, see e.g.~\cite{King:2001uz,Ross:2004qn,Antusch:2008jf}.} As in the previous section, we introduce $\SUF$-breaking fields $T$, $S$ in the ${\bf 3}$ and $\wt T$, $\wt S$ in the $\bar{\bf 3}$, which are treated as background fields from now on. We ignore $X$, $\wt X$, $Y$, and $\wt Z$; direct superpotential couplings between them and the visible sector can be forbidden e.g.~by $R$-symmetry. 

In addition to the $\mathbb{Z}_2$ symmetry of the previous section under which $T$ and $\wt T$ are odd and all other fields even, we impose a $\mathbb{Z}_4$ symmetry under which only $S$ and $\wt S$ are charged with charges $1$ and $2$ respectively. We assume that all these fields develop vacuum expectation values satisfying
\be
\vev{\wt T}^\dag=c\;\vev{T}\,,\qquad\vev{\wt S}^\dag=e^{i\phi}\vev{S}\,,
\ee
where $c$ is an ${\cal O}(1)$ constant. In other words, the VEVs of $T$ and $\wt T^\dag$ are aligned in flavour space, and the VEVs of $S$ and $\wt S^\dag$ differ only by a phase; as we have argued in the previous section, this can easily be realized dynamically.

A further crucial assumption is that $|\vev{T}|$ and $|\vev{\wt T}|$ are of the order of some cutoff scale $\Lambda$, while $|\vev{S}|$ and $|\vev{\wt S}|$ are suppressed with respect to $\Lambda$ by a factor $\epsilon\sim{\cal O}(0.05)$. Without loss of generality, we can then choose a basis where
\be\label{topyuk}
\vev{T}=\left(\begin{array}{c} 0 \\ 0 \\ v\end{array}\right)\,,\quad\vev{S}=\left(\begin{array}{c} 0 \\ u \\ w\end{array}\right)\,,
\ee 
where $v/\Lambda\sim{\cal O}(1)$, $u/\Lambda\sim{\cal O}(\epsilon)$, and $w/\Lambda\sim{\cal O}(\epsilon)$. This shows explicitly that $\SUF$ is broken to $\SU{2}_{\rm F}$ at a scale $|\vev{T}|\sim\Lambda$, and $\SU{2}_{\rm F}$ is subsequently completely broken at a lower scale $|\vev{S}|\sim\epsilon\Lambda$.

Since the cutoff $\Lambda$ is of the order of the messenger scale, a potentially relevant source for soft masses are quartic terms coupling the hidden sector to the visible sector in the K\"ahler potential, such as $|X|^2\, |Q|^2/\Lambda^2$. These terms will in fact be induced at the two-loop level by the usual gauge mediation diagrams. They are subdominant with respect to the one-loop gauge messenger contribution to the soft masses, provided that they are not already generated at one loop or at the tree level. This should be ensured by appropriate symmetries of the UV completion, analogous to messenger parity in ordinary gauge mediation.

The top Yukawa coupling is generated by the superpotential operator
\be\label{opsyt}
O_1=\lambda_1^u\frac{1}{\Lambda^2}\; {\wt T}_i U_i{\wt T}_j Q_j H_u
\ee
after replacing $\wt{T}$ with its VEV. Further contributions to the up-type quark Yukawa matrix, suppressed by powers of $\epsilon$, come from the superpotential operators
\be
\begin{split}\label{opsyuk}
O_2=\;&\lambda_2^u\,\frac{1}{\Lambda^2}(\wt S U)(\wt S Q)H_u\,,\\
O_3=\;&\lambda_3^u\,\frac{1}{\Lambda^3}(\wt S S)(S U Q)H_u\,,\\
O_4=\;&\lambda_4^u\,\frac{1}{\Lambda^4}(\wt S T)(\wt T Q)(\wt S U)H_u\,,\\
O_5=\;&\lambda_5^u\,\frac{1}{\Lambda^4}(\wt S T)(\wt S Q)(\wt T U)H_u\,,\\
O_6=\;&\lambda_6^u\,\frac{1}{\Lambda^5}(T S Q)(\wt S U) (\wt T S)H_u\,,\\
O_7=\;&\lambda_7^u\,\frac{1}{\Lambda^5}(T S U)(\wt S Q) (\wt T S)H_u\,,\\
O_8=\;&\lambda_8^u\,\frac{1}{\Lambda^5}(T S Q)(\wt T U) (\wt S S)H_u\,,\\
O_9=\;&\lambda_{9}^u\,\frac{1}{\Lambda^5}(T S U)(\wt T Q) (\wt S S)H_u\,,\\
O_{10}=\;&\lambda_{10}^u\,\frac{1}{\Lambda^5}(T U Q)(\wt T S)(\wt S S)H_u\,,\\
O_{11}=\;&\lambda_{11}^u\,\frac{1}{\Lambda^8}(S \wt T)(S \wt T)(T S Q)(T S U)H_u\,,
\end{split}
\ee
Here we have dropped the $\SUF$ indices in favour of the shorthand notation $(\wt A B)\equiv\wt A_i B_i$ and $(ABC)\equiv\epsilon_{ijk} A_i B_j C_k$. These operators give the leading-order contributions to the matrix elements of $Y_u$. The contributions from all other operators allowed by $\SUF$ and $\mathbb{Z}_2\times\mathbb{Z}_4$ are of higher order in $\epsilon$ (except that any of the $O_i$ can be multiplied with a function of $(\wt T T)$ which is ${\cal O}(1)$, but which can be absorbed in the $\lambda^u_i$ couplings). The resulting Yukawa matrix is of the form
\be\label{YU}
Y_u\sim\left(\begin{array}{ccc}
\epsilon^4 & \epsilon^3 & \epsilon^3\\
\epsilon^3 & \epsilon^2 & \epsilon^2 \\
\epsilon^3 & \epsilon^2 & 1 \\
             \end{array}\right)\,.
\ee

For the down quark sector we can write down the equivalent terms with couplings $\lambda_{i}^d$. A realistic Yukawa hierarchy requires $\lambda^d_{1}$ to be accidentally somewhat small, $\lambda^d_{1}\sim{\cal O}(\epsilon)$. The Yukawa matrix becomes 
\be\label{YD}
Y_d\sim\left(\begin{array}{ccc}
\epsilon^4 & \epsilon^3 & \epsilon^3\\
\epsilon^3 & \epsilon^2 & \epsilon^2 \\
\epsilon^3 & \epsilon^2 & \epsilon \\
             \end{array}\right)\,.
\ee

The resulting Yukawa couplings are
\be\begin{split}
(y_u,\,y_c,\,y_t)&\;\sim(\epsilon^4,\,\epsilon^2,\,1)\\
(y_d,\,y_s,\,y_b)&\;\sim(\epsilon^4,\,\epsilon^2,\,\epsilon)
\end{split}
\ee
and the CKM matrix is
\be
V_{\rm CKM}\sim\left(\begin{array}{ccc} 1 & \epsilon & \epsilon^2 \\ \epsilon & 1 & \epsilon \\ \epsilon^2 & \epsilon & 1\end{array}\right)\,.
\ee
The exact values for of the CKM angles and Yukawa couplings can be written, in an expansion in $\epsilon$, as functions of $c$, $v/\Lambda$, $u/\Lambda$, $w/\Lambda$, and of the couplings $\lambda_i^u$ and $\lambda_i^d$.
For $\epsilon\approx 0.05$ this roughly reproduces the observed flavour hierarchy, although some observables such as $V_{us}$ are slightly too suppressed, which needs to be compensated by the unknown  coefficients.\footnote{A similar pattern was advocated e.g.~in \cite{Yanagida:1998jk,Hebecker:2002re,Dreiner:2006xw,Nomura:2007ap}.}  We have checked that it is nevertheless possible to fit all quark masses and mixings with ${\cal O}(1)$ coefficients (see Appendix B for details). If the scale $\Lambda$, associated with the up-type quarks, is taken different from the scale $\bar\Lambda$, at which the operators of Eqns.~\eqref{opsyt} and \eqref{opsyuk} for the down-type quarks are generated, one could improve the fit even further by having two distinct expansion parameters $\epsilon$ and $\bar\epsilon$ \cite{King:2001uz}, but we will not do so here.

In the flavour basis of Eqns.~\eqref{YU} and \eqref{YD} the gauge messenger contributions to the squark soft masses Eq.~\eqref{gmmsoft} are diagonal but non-universal. Therefore, in the super-CKM basis where the Yukawa matrices are diagonal, off-diagonal entries in the squark mass matrices will appear, inducing potentially dangerous FCNCs. In the next section we will investigate the constraints and possible observable consequences following from this.

\subsection{Flavour violation}\label{sec:FCNC}

So far we have ignored the subleading off-diagonal squark masses which are also generated by gauge messengers. As we have already stated, Eq.~\eqref{gmmsoft} holds only in the flavour basis of Eq.~\eqref{YU}. Rotating to the super-CKM basis (in which we denote the soft matrices by $\hat m_{Q,U,D}^2$) induces
\be\begin{split} \label{deltamQsq}
\left(\hat m_{Q}^2\right)_{23}&=\left(\hat m_{Q}^2\right)_{32}^*=-\frac{\gF^2}{16\pi^2}\,\LamF^2\cdot\frac{3}{2}\;V_{ts}+{\cal O}(\epsilon^2)\,,\\
\left(\hat m_{Q}^2\right)_{13}&=\left(\hat m_{Q}^2\right)_{31}^*=-\frac{\gF^2}{16\pi^2}\,\LamF^2\cdot\frac{3}{2}\;V_{td}+{\cal O}(\epsilon^3)\,,\\
\left(\hat m_{Q}^2\right)_{12}&=\left(\hat m_{Q}^2\right)_{21}^*=-\frac{\gF^2}{16\pi^2}\,\LamF^2\cdot\frac{3}{2}\;V_{td}^* V_{ts}+{\cal O}(\epsilon^4)\,.\\
\end{split}
\ee
To leading order, the off-diagonal terms in $\hat m_Q^2$ can be expressed in terms of CKM matrix entries. This is because $\hat m_Q^2$ is given by
\be
\hat m_Q^2=V_D^\dag m_Q^2 V_D
\ee
and $V_{\rm CKM}=V_U^\dag V_D$ with $V_U$ differing from the unit matrix only by terms ${\cal O}(\epsilon^2)$. Therefore, to leading order the CKM matrix and the left-handed down-type mixing matrix $V_D$ coincide. Together with Eq.~\eqref{deltamsq} this immediately leads to  Eqs.~\eqref{deltamQsq}.

Eqs.~\eqref{deltamQsq} can in principle be used to derive simple analytic estimates for the mass insertions $\delta_{ij,{\rm (LL)}}^d$, in terms of the ratio $m^2_{{\rm GM}}/m^2_{Q,\chi {\rm M}}$ and the CKM matrix elements. Here $m^2_{Q,\chi{\rm M}}$ is the chiral messenger contribution to $m_Q^2$, given e.g.~by Eqs.~\eqref{mgmmasses} for minimal gauge mediation, $m_{\rm GM}^2$ is the gauge messenger contribution, and $\delta_{ij,{\rm (LL)}}^d$ is as usual defined by
\be
\delta_{ij,{\rm (LL)}}^d=\frac{(M_{\tilde d}^2)_{ij}}{\sqrt{(M_{\tilde d}^2)_{ii}(M_{\tilde d}^2)_{jj}}}\,.
\ee
with $M_{\tilde d}^2$ the down-type squark mass matrix. However, such expressions are of limited use because they only hold at the mediation scale, and the squark masses change substantially during renormalization group running. This is especially important for the case which we are most interested in, namely, the case of small third-generation squark masses at the electroweak scale. Therefore, one cannot directly compare the mass insertions obtained from Eqs.~\eqref{deltamQsq} with the experimental constraints. Eqs.~\eqref{deltamQsq} are nevertheless instructive, since we can directly read off the order of magnitude of suppression for the $\delta_{ij,{\rm (LL)}}^d$.

The off-diagonal entries of $\hat m_D^2$ in the SCKM basis are not directly related to any CKM matrix entries. They can however be parameterized as 
\be\begin{split} 
\left(\hat m_{D}^2\right)_{23}&=\left(\hat m_{D}^2\right)_{32}^*=-\frac{\gF^2}{16\pi^2}\,\LamF^2\cdot\frac{3}{2}\;\eta_{23}\;\epsilon+{\cal O}(\epsilon^2)\,,\\
\left(\hat m_{D}^2\right)_{13}&=\left(\hat m_{D}^2\right)_{31}^*=-\frac{\gF^2}{16\pi^2}\,\LamF^2\cdot\frac{3}{2}\;\eta_{13}\;\epsilon^2+{\cal O}(\epsilon^3)\,,\\
\left(\hat m_{D}^2\right)_{12}&=\left(\hat m_{D}^2\right)_{21}^*=-\frac{\gF^2}{16\pi^2}\,\LamF^2\cdot\frac{3}{2}\;\eta_{12}\;\epsilon^3+{\cal O}(\epsilon^4)\,,
\end{split}
\ee
where the $\eta_{ij}$ are some ${\cal O}(1)$ constants depending on the $\lambda_i^d$ coefficients of the flavour model. 

The off-diagonal entries of $\hat m_U^2$ induced by the CKM rotation are highly suppressed. Hence, even though the constraints on flavour violation for up-type squarks are becoming increasingly competitive, we will ignore them from now on.

Another source of off-diagonal soft masses are the subleading corrections to Eq.~\eqref{deltamsq} due to the non-vanishing $S$ and $\wt S$ VEVs. These are obtained by applying Eqns.~\eqref{ZT} and \eqref{mQIsq} to the full set of $\SUF$ breaking VEVs and expanding in $\epsilon$, as detailed in Appendix A. The result is
\be\label{deltamsq23}
\left(\delta m_{Q,U,D}^2\right)_{23}=\left(\delta m_{Q,U,D}^2\right)_{32}^*=-\frac{\gF^2}{16\pi^2}\,\LamF^2\,\eta\left(\frac{3}{2}+\frac{1}{2}\log\left(\frac{3}{4}\epsilon^6\right)\right)\epsilon^2\,.
\ee
Here $\eta=w/u$, where $u$ and $w$ are the VEVs of the flavour-breaking field $S$. To leading order in $\epsilon$, these contributions can simply be added to $\hat m_{Q,U,D}^2$ in the SCKM basis. 

The off-diagonal elements in the squark mass matrix may lead to sizeable new physics contributions to flavour physics observables. The most stringent constraints come from $K$-$\bar K$ mixing. In the mass insertion approximation, $\delta^d_{12({\rm RR,LL})}$ and the double mass insertion $\delta^d_{13({\rm RR,LL})}\delta^d_{23({\rm RR,LL})}$ contribute at the same order in $\epsilon$. If $\epsilon$ is ${\cal O}(0.05)$, then the effect is estimated to be ${\cal O}(10^{-4})$, which is a very interesting region for squark and gluino masses in the range of $1$--$2$ TeV, bordering on being excluded. Constraints on natural SUSY from $K$-$\bar K$ mixing were discussed in \cite{Giudice:2008uk} and more recently in \cite{Mescia:2012fg}.

To investigate the impact of flavour physics constraints on our model quantitatively, we have sampled the parameter space of our flavour model using a simple Markov Chain Monte Carlo method. Taking the vacuum expectation values of $v$, $u$ and $w$ as well as the constants $c$, $\lambda^u_i$ and $\lambda^d_i$ as free parameters, their values were determined such that the Yukawa couplings and CKM data were reproduced as measured. Note that there is a large ambiguity in doing so, as there are many more free parameters than observables. Restricting $|\lambda^{u,d}_i|$, $|c|$ and $|v|/\Lambda$ to be ${\cal O}(1)$, and $|u|/\Lambda$ and $|w|/\Lambda$ to be ${\cal O}(\epsilon)$, we obtain a distribution of valid parameter points which are then used to calculate the effects on flavour observables which we expect in this model. More details on our method are given in Appendix B.

For the spectrum of Section \ref{sec:softterms}, we have checked the resulting model predictions for $\epsK$, $\Delta m_K$, $\Delta m_{B_d}$ and ${\rm BR}(b\into X_s\gamma)$. The most severe constraints come from $\epsK$, since this observable can be calculated quite precisely. For all other observables our model reproduces the measured values fully within the theoretical uncertainty.

\begin{figure}
\centering
\includegraphics[width=.48\textwidth]{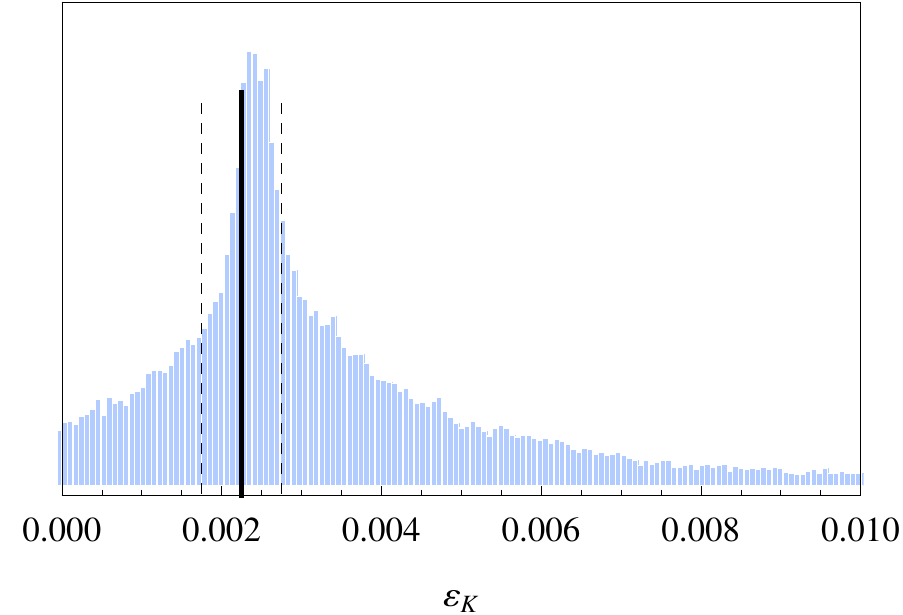}
\caption{Range of $\epsK$ in the flavour model of Section \ref{sec:yukckm} as computed with {\tt SUSY\_FLAVOR 2} \cite{Crivellin:2012jv}, using the particle spectrum of Section \ref{sec:softterms}. The measured value of $\epsK=2.23\times 10^{-3}$ is indicated, together with an estimated theory uncertainty interval of $\pm 0.5\times 10^{-3}$.}\label{fig:epsK}
\end{figure}

The $\epsK$ distribution is shown in Fig.~\ref{fig:epsK}. Significant deviations from the measured value $\epsK=2.23\times 10^{-3}$ are evidently possible in our model. However, the width of the distribution is not much larger than the theoretical uncertainty of about $20\%$ (the experimental error of $\pm 0.01\times 10^{-3}$ is insignificant by comparison), and it is peaked near the observed value. This indicates that, at present, our model is compatible with flavour precision experiments without major fine-tuning. Still one should generically expect some deviation from the Standard Model value, which will become more significant as the reliability of the theoretical predictions improves.

\section{Discussion and conclusions}\label{sec:conclusions}

In this paper, we have used a gauged, non-supersymmetrically broken $\SUF$ quark flavour symmetry to give new contributions to the soft term spectrum of gauge-mediated supersymmetry breaking. SUSY breaking is aligned with $\SUF\into\SU{2}_{\rm F}$ breaking, which is responsible for generating the third-generation Yukawa couplings, causing the massive gauge supermultiplets to induce a flavour non-universal squark soft mass. This contribution is negative, appears at one loop in the $\SUF$ coupling, and affects mainly the third generation. Together with the usual positive and flavour-universal soft masses from ordinary gauge mediation, one may obtain sub-TeV stop and sbottom squarks while the first- and second-generation squarks remain above the present LHC exclusion limits. We have shown that the required alignment of SUSY breaking and flavour breaking can be realized dynamically. Moreover, the induced off-diagonal squark masses can be calculated when the flavour breaking model is 
specified, and one may compare the resulting flavour-violating effects with the experimental constraints on FCNCs.

A soft term pattern such as the one we have advocated could leave its imprint on three very different classes of experimental searches: Standard SUSY searches for first-generation squarks and gluinos which undergo cascade decays into jets and missing energy; dedicated searches for stop and sbottom squarks; and searches for new physics contributions to FCNCs. This is of course very unusual for a gauge-mediated model, the hallmark of standard gauge mediation being its flavour universality. In our model mild FCNCs may be introduced in a controlled fashion and of course vanish entirely in the limit $\gF\rightarrow 0$. Our analysis shows that flavour gauge messengers can change the superpartner spectrum significantly with respect to simple gauge-mediated models.

In our study, the two-loop contributions to scalar soft masses from chiral messengers and the one-loop contributions from gauge messengers were taken to be comparable. This is to some extent an arbitrary choice: A priori the $\SUF$ gauge coupling $\gF$ could also be much smaller (in which case there would be no noticeable effect), or much larger (but then the spectrum would suffer from tachyons). Note however that this choice does not represent a fine-tuning, since we are not cancelling two large quantities against each other to produce a tiny outcome; we are merely choosing the two effects to be of the same order of magnitude. Note also that a somewhat small $\gF$ is consistent with the fact that $\SUF$ is asymptotically non-free, because the number of matter fields charged under $\SUF$ is quite large (the exact number being model-dependent). Searches for stops \cite{ATLAS3rd, CMS3rd} put a lower bound on the combination $\gF^2\LamF^2$.

On the other hand, the light stop squarks resulting from flavour gauge messengers cannot be argued to improve the supersymmetric little hierarchy problem (thereby providing an interesting counterexample to the claim that lighter stops are always more natural). Since a small stop mass in our model is always the sum of two relatively large opposite-sign contributions, the sensitivity of the electroweak scale to the fundamental parameters is not reduced significantly. At most one can argue that within the MSSM a 125 GeV Higgs is somewhat easier to obtain, since the required $A$-terms need no longer be extremely large.

If one insists on unification, the flavour symmetry should be extended also to the lepton sector. While this is straightforward from the model-building point of view, an immediate and undesired consequence would be large tachyonic contributions also to the stau masses. The right-handed stau is typically the LSP in gauge-mediated models without gauge messengers (not counting the gravitino). Requiring that its mass remains positive then would limit the maximal gauge messenger effect (more precisely, there would be a lower bound on $\gF^2\LamF^2$ from searches for taus with MET \cite{CMS:2013jfa,ATLAS:2013ama}), and therefore also the amount by which the stop masses can be reduced.

It would be interesting to generalize our mechanism to other realistic patterns of flavour symmetry breaking, using a more elaborate flavour model in which the mass and CKM hierarchies are more naturally reproduced. Models based on other flavour symmetries such as $\SU{3}_{\rm F,L}\times\SU{3}_{\rm F,R}$, where $\SU{3}_{\rm F,L}$ acts only on the left-handed and $\SU{3}_{\rm F,R}$ acts only on the right-handed quarks, could also be of interest, as could more conventional abelian Froggatt-Nielsen models based on $\U{1}$ symmetries with generation-dependent charges. Note however that, in order to obtain a large effect for the third-generation squarks only, their $\U{1}$ charge would have to be large while that of the first two generations would have to be small (and near-degenerate to avoid FCNCs). Obtaining a realistic flavour hierarchy would therefore be more difficult, leading us to believe that an $\SUF$-based model such as ours may indeed be the simplest approach. Another promising direction for future 
research may be to embed the flavour-breaking mechanism into a more complete model of dynamical SUSY breaking. 

\subsection*{Acknowledgements}

The authors thank Thomas Konstandin, Ben O'Leary, Florian Staub and Alexander Westphal for correspondence and discussions about vacuum stability, and Robert Ziegler for comments on the manuscript. MM is funded by the Alexander von Humboldt Foundation. The work of AW was supported in part by the German Science Foundation (DFG) under the Collaborative Research Center (SFB) 676.

\appendix

\section*{Appendix A: One-loop scalar mass to order $\epsilon^2\log\epsilon$}

Here we give the generalization of Eq.~\eqref{deltamsq} when taking the subleading $S$ and $\wt S$ VEVs into account, to leading order in an expansion in the flavour hierarchy parameter $\epsilon$. The model is defined in Section \ref{sec:flmod}. We set $c=1$ for simplicity (the more general case is straightforward).  The gauge group is $\SUF$ and the hidden sector comprises the fields $\{Z_i\}=\{T,\wt T,S,\wt S,X,\wt X\}$, with untilded fields transforming as ${\bf 3}$ and tilded ones as $\bar{\bf 3}$. The vacuum expectation values are
\be\begin{split}
\vev{T}&=\left(\begin{array}{c} 0 \\ 0 \\ v \end{array}\right)\,,\quad \vev{S}=\left(\begin{array}{c} 0 \\ u \\ w \end{array}\right)\,,\quad\vev{X}=\left(\begin{array}{c} 0 \\ 0 \\ F_X\,\theta^2 \end{array}\right)\,,\\
\quad \vev{\wt T}&=\left(\begin{array}{ccc} 0 & 0 & v^* \end{array}\right)\,,\quad \vev{\wt S}=e^{i\phi}\left(\begin{array}{ccc} 0 & u^* &  w^* \end{array}\right)\,,\quad \vev{\wt X}=\left(\begin{array}{ccc} 0 & 0 & F_X^*\,\theta^2 \end{array}\right)\,.
\end{split}
\ee
We define
\be
M=\gF\,|v|\,,\quad\epsilon=|u/v|\,,\quad\eta=w/u\,.
\ee
As explained in the main text, we should have $\epsilon\sim{\cal O}(0.05)$ and $|\eta|\sim{\cal O}(1)$. Five of the gauge boson mass eigenstates will then acquire supersymmetric masses ${\cal O}(M^2)$, and the remaining three will be somewhat lighter with masses ${\cal O}(\epsilon^2 M^2)$. We assume $F_X< \epsilon^2 M^2$, since the effective K\"ahler potenial approach is only valid in the limit of small SUSY breaking.

To leading order in $\epsilon$ and $F$, from Eqns.~\eqref{ZT} and \eqref{mQIsq} one obtains the following one-loop soft mass squared for any $Q_I$ which transforms as a ${\bf 3}$ under $\SUF$:
\be\begin{split}\label{oneloopmsq}
m_{Q_I}^2=&\;-\frac{\gF^2}{16\pi^2}\frac{|F_X|^2}{|v|^2}\;\left\{\;\left(\begin{array}{ccc}\frac{7}{6} & 0 & 0 \\ 0 & \frac{7}{6} & 0 \\ 0 & 0 & \frac{8}{3} \end{array}\right)\right.\\
&\qquad +
\left.\left(\begin{array}{ccc}
\frac{1}{3}|\eta|^2-\frac{1}{6}-\frac{1}{4}\log\left(\frac{3}{4}\epsilon^2\right) & 0 & 0\\
0 & \frac{1}{3}|\eta|^2-\frac{2}{3}+\frac{1}{4}\log\left(\frac{3}{4}\epsilon^2\right) & \eta\left(\frac{3}{2}+\frac{1}{2}\log\left(\frac{3}{4}\epsilon^6\right)\right) \\
0 &  \eta^*\left(\frac{3}{2}+\frac{1}{2}\log\left(\frac{3}{4}\epsilon^6\right)\right) & -\frac{8}{3}|\eta|^2-\frac{7}{6}
\end{array}\right)\epsilon^2\;\right\}\\
&\;+\;{\cal O}\left(\epsilon^3,|F_X|^3/|v|^4\right)\,.
\end{split}
\ee
In fact, we find that $(m_{Q_I}^2)_{12}=(m_{Q_I}^2)_{13}=0$ to all orders in $\epsilon$, so $(m_{Q_I}^2)_{23}$ and $(m_{Q_I}^2)_{32}$ are the only off-diagonal mass matrix elements in this flavour basis.

In the more general case that $c\neq 0$ and that there are more fields with lowest- or highest-component VEVs aligned with $\vev{T}$, one should replace $|F_X|^2\into\sum_i |F_i|^2$ and $|v|^2\into\sum_i |v_i|^2$ in Eq.~\eqref{oneloopmsq}. 

The above tachyonic contributions to the scalar masses are the most important effect of the gauge messengers in our model. Trilinear $A$-terms are also induced at one loop, but since the $A$-parameters have mass dimension one, these are clearly subdominant compared to the scalar masses. All other visible sector soft terms are associated with fields which do not carry $\SUF$ charges, and are therefore only generated at even higher loop order.

\section*{Appendix B: Details on the flavour model scan}

To estimate the flavour violation effects in our model, we have followed a procedure which we now briefly describe. Our flavour model is defined by the operators in Eqns.~\eqref{topyuk} and \eqref{opsyuk}, and the same operators with $(U_n,\lambda^u_i)$ replaced by $(D_n,\lambda^d_i)$. The coefficients $\lambda^{u,d}_i$ are a priori complex and of order one (except for $\lambda^d_1$ which we take to be ${\cal O}(\epsilon)$), but otherwise anarchic. Together with the constant $c$, the VEVs $v,u,w$, the scale $\Lambda$, and the relative phase $\phi$ between $\vev S$ and $\vev{\wt S^\dag}$ this amounts to 54 real parameters. Even when taking into account that several of these are spurious or of subleading influence, there is still a large redundancy when fitting only ten observables (six quark masses, three CKM angles, and one CKM phase). We therefore use a Markov Chain Monte Carlo method, which is well suited for sampling models with a large number of free parameters.

The parameters which are actually relevant are $\{a_\alpha\}=\{\lambda^d_i,\,\lambda^u_1,\,\lambda^u_2,\,\lambda^u_4,\,\lambda^u_5,\,c,\,u,\,w,\,\phi,\,\epsilon\}$. Here $\lambda_1^d$ is normalized to $\epsilon$ and $u$ and $w$ are normalized to $\epsilon\Lambda$, such that all $a_\alpha$ except $\epsilon$ are ${\cal O}(1)$ and dimensionless. We take them to be in the range $\frac{1}{3}\leq|a_\alpha|\leq 3$ with arbitrary phases, except for $\phi\in [0,\;2\pi)$ and $\epsilon\in [0,\;0.2]$. The $a_\alpha$ are then fed into a Metropolis-Hastings MCMC code which tries to fit $y_u$, $y_c$, $y_b$, $y_s$, $y_d$, $|V_{us}|$, $|V_{ub}|$, $|V_{cb}|$, and the Jarlskog invariant $J_{\rm CKM}$ at the mediation scale $M$, for a given SUSY spectrum.\footnote{In this model, the parameters affecting the up quark Yukawa coupling $y_u$ turn out to be severely underconstrained and quite irrelevant for the flavour observables, so we can omit them from the fit.} For data points which properly reproduce the observables, the off-
diagonal corrections to the squark mass matrices are calculated at $M$ (these corrections have negligible influence on the values of the fitted observables, so there is no need for iterating the procedure). The spectrum is then evolved with {\tt SPheno} to the electroweak scale, where the result is passed to {\tt SUSY\_FLAVOR 2} \cite{Crivellin:2012jv}.

\end{document}